\documentclass[12pt]{article}
\usepackage{mydef}
\usepackage{authblk}
\usepackage{epsfig}

\usepackage[hide]{ed}

\begin{document}

\title{High-Order Splitting Methods for Forward PDEs and PIDEs.\thanks{Opinions expressed in this paper are those of the author, and do not necessarily reflect the view of Numerix or NYU.}}
\author{Andrey Itkin}

\affil{\small Numerix LLC, \\125 Park Avenue, 21th Floor, New York, NY 10017, USA \\
and \\
Polytechnic School of Engineering, New York University, \\
6 Metro Tech Center, RH 517E, Brooklyn NY 11201, USA}

\date{}

\maketitle
\vspace{-0.3in}
\centerline{\footnotesize Submitted to the International Journal of Theoretical and Applied Finance}
\vspace{0.1in}

\begin{abstract}
This paper is dedicated to the construction of high-order (in both space and time) finite-difference schemes for both forward and backward PDEs and PIDEs, such that option prices obtained by solving both the forward and backward equations are consistent. This approach is partly inspired by \cite{AndreassenHuge2011} who reported a pair of consistent finite-difference schemes of first-order approximation in time for an uncorrelated local stochastic volatility model.  We extend their approach by constructing schemes that are second-order in both space and time and that apply to models with jumps and discrete dividends.  Taking correlation into account in our approach is also not an issue.
\end{abstract}

\vspace{0.5in}

\section{Introduction}
\label{introduction}
Constructing consistent numerical methods for a given model (e.g., option pricing using various numerical schemes to compute the price with the same tolerance for all methods) is a known problem. Recently, this problem has come under attention for local stochastic volatility (LSV) models. For instance, \cite{AndreassenHuge2011} present a numerical solution that achieves full consistency between calibration, option pricing using finite-difference solutions and Monte Carlo simulation. The method is based on a fully implicit finite-difference scheme of first-order in time, which was motivated by the fact that Monte Carlo simulation, even with the use of the Milstein scheme, is also of first-order in time, see \cite{PGbook} and references therein. However, in many practical applications we only need consistency between the solutions obtained with the forward and backward Kolmogorov equations.
We recall here that the forward equation is useful for calibration as it allows computation of the option smile in one sweep,\footnote{In other words, option prices for multiple strikes given the spot price could be computed by solving just one forward equation.} while the backward equation is useful for pricing as it allows computation of option prices with various initial spot values for a given strike in one sweep.
In this case consistent schemes of higher-order approximation in both space and time would be desirable. It is also interesting to extend such an approach to jump-diffusion models, thus adding jumps to the LSV framework.

As mentioned in \cite{AndreassenHuge2011}, if special care is not taken, then  the different numerical schemes will generally be fully consistent with each other only in the limit when the number of Fourier steps for FFT, the number of time and spatial steps in the finite-difference grid, and the number of time steps in the Monte Carlo all tend to infinity and the numerical schemes converge to the continuous time and space solution. Furthermore, setting appropriate boundary conditions to guarantee this consistency could be a nontrivial task.

These problems, however, could be eliminated if one considers a discretized Kolmogorov equation from the beginning. In other words, this is equivalent to the discrete Markov chain defined at space states corresponding to some finite-difference grid $\mathbf{G}$. Consider first only the stochastic processes with no jumps. It is known (see, e.g., \cite{Goodman2004}) that the forward Kolmogorov equation for the density of the underlying process $X(t)$ can be written in the form
\begin{equation} \label{FKE}
\frac{ \partial P}{ \partial T}  =   P \mathcal{A},
\end{equation}
\noindent where $ P({\bf s},t)$ is a discrete density function (i.e., a vector of probabilities that $X(t)$ is in the corresponding state ${\bf s}$ at time $t$), $T$ is the expiration time, and the generator $\mathcal{A}$ has a discrete representation on $\mathbf{G}$ as a matrix of transition probabilities between the states. For a given interval $ [t, t + \Delta t] $ where the generator $ \mathcal{A} $ does not depend on time,
the solution of the forward equation is
\begin{equation} \label{FKEsol}
P({\bf s},t+\Delta t) = P({\bf s},t)e^{\Delta t \mathcal{A}}.
\end{equation}
Therefore, to compute the nondiscounted option price $C = E_Q[V(X(T))]$, where $V(X(t))$ is the payoff at time $t$ and $E_Q$ is an expectation under the risk-neutral measure $Q$, we start with a given $P(0)$, evaluate $P(T)$ by solving \eqref{FKE},\footnote{If $\mathcal{A} = \mathcal{A}(t)$, we solve in multiple steps in time $\Delta t$ by using, for instance, a piecewise linear approximation of $\mathcal{A}$ and the solution in \eqref{FKEsol}.} and finally evaluate $P({\bf s},T) V({\bf s},T)$.

On the other hand, one can start with a backward Kolmogorov equation for nondiscounted option price
\begin{equation} \label{BKE}
\fp{V}{t} + \mathcal{A} V  = 0,
\end{equation}
\noindent which by change of variables $\tau = T-t$ could be transformed to
\begin{equation} \label{BKE1}
\frac{ \partial V}{ \partial \tau}  =   \mathcal{B} V.
\end{equation}
It is also well-known that $\mathcal{B} = \mathcal{A}^T$; see, e.g., \cite{Goodman2004}.

To the best of our knowledge, Alex Lipton was the first to propose using this duality to construct the consistent schemes for forward and backward equations. The idea was described in a Bankers Trust paper somewhere between 1997 and 1999, and has been known since then as ``Lipton's trick''. The method was then alluded to in \cite{Lipton2001, Lipton2002}. Later, the same idea was rediscovered in \cite{AndreassenHuge2011} and used to construct a consistent implicit finite-difference scheme that is first-order in time. See \cite{AA2000} for the forward and backward equations.

Based on the above work, constructing the transposed discrete operator $\mathcal{A}^T$ on a grid $\mathbf{G}$ is fairly straightforward. However, extending this idea  to modern finite-difference schemes of higher-order approximation (see, e.g., \cite{HoutWelfert2007} and references therein) is a greater obstacle.  This obstacle is especially challenging because these schemes by nature do multiple fractional steps to accomplish the final solution, and because an explicit form of the generator $\mathcal{A}$ that is obtained by applying all the steps is not obvious. Finally, including jumps in consideration makes this problem even harder.

Therefore, in the present paper we focus on the construction of a consistent finite-difference scheme for the forward and backward equations. For the backward equation, we rely on the schemes described in \cite{HoutWelfert2007}. Our goal here is to construct a consistent forward counterpart.\footnote{It is worth mentioning that the consistent forward scheme for the Craig-Sneyd scheme was implemented in 2009 by Michael Konikov as a part of the Numerix library.} Our main results in this paper are: i) explicit splitting schemes that cover the forward equations and that are counterparts of the backward schemes considered in \cite{HoutWelfert2007}, ii) extension of these schemes in cases when the underlying stock pays discrete dividends, and iii) analysis of the robustness of these schemes as a function of the parameter $\theta$ of the schemes.

The rest of the paper is organized as follows. In Section \ref{lsv}, we describe an LSV model with jumps and express the pricing PIDE as a sum of a diffusion operator and a jump operator.  In Section \ref{splitting}, we describe a splitting method for solving the PIDE, where we separate the diffusion operator and the jump operator into separate fractional steps.  In Section \ref{fdscheme_diffusion}, we look at the diffusion part, consider some backward schemes that were discussed in \cite{HoutFoulon2010} and develop their consistent forward counterparts. In section \ref{forwardscheme_jumps}, we discuss how this approach could be extended to the jump component.  In Section \ref{numerical_results}, we describe various issues that have to be addressed in implementing an approach that is consistent for both forward and backward schemes.  We also present and discuss the results of some numerical experiments.


\section{LSV Model with Jumps} \label{lsv}
To avoid uncertainty, let us look at the problem of pricing equity options written on a single stock. This specification does not cause any loss of generality, but it makes the description more practical. We assume that an underlying asset (stock) price $S_t$ is driven by an exponential of a L{\'e}vy process:
\begin{equation} \label{Levy}
S_t = S_0 \exp (L_t), \quad 0 \le t \le T,
\end{equation}
\noindent where $t$ is time, $T$ is the option expiration, $S_0 = S_t|_{t=0}$, and $L = \{L_t\}_{0 \le t \le T}$ is the L{\'e}vy process with a nonzero Brownian (diffusion) part. Under the pricing measure, $L_t$ is given by
\begin{equation} \label{Lt}
L_t = \gamma t  + \sigma W_t + Y_t, \qquad \gamma \in \mathbb{R}, \quad \sigma > 0,
\end{equation}
\noindent with L{\'e}vy triplet $(\gamma, \sigma, \nu$), where $W_t$ is a standard Brownian motion on $0 \le t \le T$ and $Y_t$ is a pure jump process.

We consider this process under the pricing measure, and therefore $e^{-(r-q) t} S_t$  is a martingale, where $r$ is the interest rate and $q$ is a continuous dividend. This allows us to express $\gamma$ as (\cite{Eberlein2009})
\[
\gamma = r - q - \frac{\sigma^2}{2} - \int_\mathbb{R} \left(e^x -1 -x {\bf 1}_{|x| < 1}\right)\nu(dx),
\]
\noindent where $\nu(dx)$ is a L{\'e}vy measure which satisfies
\[ \int_{|x| > 1}e^x \nu(dx) < \infty.  \]

Let us leave $\nu(dx)$ unspecified at this time because we are open to consider all types of jumps, including those with finite and infinite variation and finite and infinite activity.\footnote{We recall that a standard Brownian motion already has paths of infinite variation. Therefore, the L{\'e}vy process in \eqref{Lt} has infinite variation since it contains a continuous martingale component. However, here we refer to the infinite variation that comes from the jumps.}

Next, we extend this setup by assuming that $\sigma$ is a combination of a local volatility $\phi(S_t,t)$ and a stochastic volatility $\sqrt{v_t}$.  That is, we take $\sigma = \sigma_t \equiv \phi(S_t,t) \sqrt{v_t}$, where $v_t$ is the stochastic variance. The latter follows
\begin{align} \label{vSDE}
    d v_t &=  \kappa(v_{\infty} - v_t) dt + \xi v_t^\beta dZ_t, \\
    \langle dW_t, dZ_t\rangle &= \rho dt. \nonumber
\end{align}
Here $\kappa$, $v_{\infty}$, and $\xi$ are the mean-reversion speed, the long-term run (mean-reversion level), and the volatility of volatility, respectively, $Z_t$ is a Brownian motion that correlates with $W_t$, and $\rho$ is the correlation coefficient. The parameter $\beta$ determines a mean-reverting CEV process for $v_t$ and is assumed to be a calibrated parameter of the model, such that $\beta \geq 0$.\footnote{As instantaneous variance should not be a martingale, the upper boundary of $\beta$ could be extended to infinity.} It is know that $\beta < 1$ produces the so-called leverage effect, commonly observed in equity markets, where the volatility of a stock increases as its price falls. Conversely, in commodity markets, the so-called inverse leverage effect could be observed whereby the volatility of the price of a commodity tends to increase as its price increases.

We further assume for simplicity that $\kappa, v_{\infty}$, $\xi$, and $\rho$ are constants. However, this assumption could easily be relaxed to take into account time-dependent coefficients.

To price options written on the underlying process $S_t$, we want to derive a PIDE that describes the time evolution of the European option prices $C(x,v,t)$, where  $x \equiv \log (S_t/S_0)$ and $v = v_t$. Using a standard martingale approach, or creating a self-financing portfolio, one can derive the corresponding backward PIDE (\cite{ContTankov, Lewis:2000})
\begin{multline} \label{PIDE}
r C(x,v,t) = \fp{C(x,v,t)}{t} + \left(r-\frac{1}{2}v \right) \fp{C(x,v,t)}{x} + \frac{1}{2}v \sop{C(x,v,t)}{x} \\
+ \kappa\left(v_{\infty} - v\right) \fp{C(x,v,t)}{v} + \frac{1}{2}\xi^2 v^{2\beta}\sop{C(x,v,t)}{v} + \rho \xi v^{\beta+1} \phi(x,t) \cp{C(x,v,t)}{x}{v}
\\
+  \int_\mathbb{R}\left[ C(x+y,v,t) - C(x,v,t) - (e^y-1)\fp{C(x,v,t)}{x} \right] \nu(dy)
\end{multline}
for all $(x,v,t) \in \mathbb{R}\times \mathbb{R}^+ \times (0,T)$, subject to the terminal and initial conditions
\begin{equation}
C(x,v,T) = h(x), \quad v(0) = v_0,
\end{equation}
\noindent where $v_0$ is the initial level of the instantaneous variance, $h(x)$ is the option payoff, and some boundary conditions that depend on the type of the option are applied. The solutions of this PIDE usually belong to the class of viscosity solutions (\cite{ContTankov}).

In the next sections, we will need another representation of the jump term proposed in
\cite{ItkinCarr2012Kinky}. It was shown there that the integral term could be rewritten using the following idea. As is known from quantum mechanics (\cite{OMQM}), a translation (shift) operator in $L_2$ space can be represented as
\begin{equation} \label{transform}
    \mathcal{T}_b = \exp \left( b \dfrac{\partial}{\partial x} \right),
\end{equation}
\noindent with $b$ = const, so
\[ \mathcal{T}_b f(x) = f(x+b). \]

Therefore, the integral in \eqref{PIDE} can be formally rewritten as
\begin{align} \label{intGen}
\int_\mathbb{R} \left[ C(x+y,v,t) \right. & \left. - C(x,v,t) - (e^y-1) \fp{C(x,v,t)}{x} \right] \nu(dy) =  \mathcal{J} C(x,v,t), \\
\mathcal{J} & \equiv \int_\mathbb{R}\left[
\exp \left( y \dfrac{\partial}{\partial x} \right) - 1 - (e^y-1) \fp{}{x} \right] \nu(dy). \nonumber
\end{align}

In the definition of the operator $\mathcal{J}$ (which is actually an infinitesimal generator of the jump process), the integral can be formally computed under some mild assumptions about existence and convergence if one treats the term $\partial/ \partial x$ as a constant. Therefore, the operator $\mathcal{J}$ can be considered as some generalized function of the differential operator $\partial_x$. We can also treat $\mathcal{J}$ as a pseudo-differential operator.

With allowance for this representation, the entire PIDE in \eqref{PIDE} can be rewritten in operator form as
\begin{equation} \label{oper}
\partial_\tau C(x,v,\tau) = [\mathcal{D} + \mathcal{J}]C(x,v,\tau),
\end{equation}
\noindent where $\tau = T - t$ and $\mathcal{D}$ represents a differential (parabolic) operator
\begin{equation} \label{diffOper}
\mathcal{D} \equiv - r + \left(r-\frac{1}{2}v^2 \right) \fp{}{x} + \frac{1}{2}v \sop{}{x}
+ \kappa\left(v_{\infty} - v\right) \fp{}{v} + \frac{1}{2}\xi^2 v^{2\beta}\sop{}{v} + \rho \xi v^{\beta+1} \phi(x,t) \cp{}{x}{v},
\end{equation}
which is an infinitesimal generator of diffusion.

\section{Splitting Method for the PIDE} \label{splitting}
To solve \eqref{oper}, we use a splitting method in a similar manner to \cite{Itkin2013} (see also references therein for a general description of splitting methods for solving PDEs).
Consider $f({\bf x},t)$ to be a function of independent variables ${\bf x} = x_1 ... x_k$ and $t$, and $\mathcal{L}$ to be a linear $k$-dimensional operator in ${\bf x}$-space. Based on \cite{LanserVerwer} consider the equation
\begin{equation} \label{operEq}
\fp{f({\bf x},t)}{t} = \mathcal{L} f({\bf x},t),
\end{equation}
Decomposing the total (compound)  operator $\mathcal{L}$ for problems of interest seems natural if, say, $\mathcal{L}$ can be represented as a sum of $k$ noncommuting linear operators $\mathcal{L}=\sum_{i=1}^k \mathcal{L}_i$. In this case, the operator equation \eqref{operEq} can be formally integrated via an operator exponential, i.e.,
\[ f({\bf x},t) = e^{t \mathcal{L}} f({\bf x}, 0)  = e^{ t \sum_{i=1}^k \mathcal{L}_i} f(0). \]
The latter expression could be factored into a product of operators:
\[ f({\bf x},t) = e^{t \mathcal{L}_k} \dots  e^{t \mathcal{L }_1}f({\bf x},0).\]
This equation could then be solved in $N$ steps sequentially by the following procedure:
\begin{align*}
f^{(1)}({\bf x},t) &= e^{t \mathcal{L}_1}f({\bf x}, 0), \nonumber \\
f^{(2)}({\bf x},t) &= e^{t \mathcal{L}_2}f^{(1)}({\bf x},t), \nonumber \\
& \quad \vdots \\
f^{(k)}({\bf x},t) &= e^{t \mathcal{L}_k}f^{(k-1)}({\bf x},t), \nonumber \\
f({\bf x}, t) &= f^{(k)}({\bf x},t). \nonumber
\end{align*}
This algorithm is exact (no bias) if all the operators $(\mathcal{L}_i)$ commute. If, however, they do not commute, the above factorization does not hold, and the algorithm provides only a first-order approximation in time (i.e., $O(t)$) to the exact solution.

To get the second-order splitting for non-commuting operators, Strang proposes another scheme, which in the simplest case ($k=2$) is (\cite{Strang})
\begin{equation}
f({\bf x},t) = e^{t L} f({\bf x},0) = e^{ t (L_1 + L_2)} f({\bf x},0) = e^{ \frac{t}{2} L_1 } e^{t L_2} e^{ \frac{t}{2} L_1 } f({\bf x},0) + O(t^2).
\end{equation}

For parabolic equations with constant coefficients, this composite algorithm is second-order accurate in $t$ as long as the numerical procedure that solves the corresponding equation at each splitting step is at least second-order accurate.

The above analysis, however, cannot be directly applied to our problem, because after the transformation in \eqref{transform} is done, the jump integral transforms to the nonlinear operator in \eqref{intGen}.
For {\it nonlinear} operators, the situation is more delicate. As shown in \cite{ThalhammerKoch2010}, the theoretical analysis of the nonlinear initial value problem
\[ u'(t) = F(u(t)), \qquad 0 \le t \le T \]
for a Banach-space-valued function $u:[0,T] \rightarrow X$ given an initial condition $u(0)$ could be done using the calculus of Lie derivatives. A formal linear representation of the exact solution is
\[ u(t) = \mathcal{E}_F(t,u(0)) = e^{t D_F} u(0), \qquad 0 \le t \le T, \]
\noindent where the evolution operator and Lie derivatives are given by
\begin{align*}
e^{t D_F} v &= \mathcal{E}_F(t,v), \quad e^{t D_F} G v = G(\mathcal{E}_F(t,v)), \quad 0 \le t \le T, \\
D_F v &= F(v), \quad D_F G v = G'(v) F(v)
\end{align*}
\noindent for an unbounded nonlinear operator $G: D(G) \subset X \rightarrow X$. Using this formalism, \cite{ThalhammerKoch2010} show that Strang's  second-order splitting method remains unchanged in the case of nonlinear operators.

Using this result for \eqref{oper} gives rise to the following numerical scheme for the backward equation:
\begin{equation} \label{bwSplit}
C(x,v,\tau+\Delta \tau) = e^{\frac{\Delta \tau}{2} \mathcal{D} } e^{\Delta \tau \mathcal{J}}
e^{\frac{\Delta \tau}{2} \mathcal{D} } C(x,v,\tau),
\end{equation}
\noindent which could be re-written as a fractional steps method:
\begin{align} \label{splitFin}
C^{(1)}(x,v,\tau) &= e^{\frac{\Delta \tau}{2} \mathcal{D} } C(x,v,\tau), \\
C^{(2)}(x,v,\tau) &= e^{\Delta \tau \mathcal{J}} C^{(1)}(x,v,\tau) \nonumber, \\
C(x,v,\tau+\Delta \tau) &= e^{\frac{\Delta \tau}{2} \mathcal{D} } C^{(2)}(x,v,\tau).  \nonumber
\end{align}
Thus, instead of an unsteady PIDE, we obtain one PIDE with no drift and diffusion (the second equation in \eqref{splitFin}) and two unsteady PDEs (the first and third ones in \eqref{splitFin}).

To produce a consistent discrete forward equation, we use the transposed evolution operator, which results in the scheme
\begin{equation} \label{fwSplit}
C(x,v,t+\Delta t) = e^{\frac{\Delta \tau}{2} \mathcal{D}^T } e^{\Delta \tau \mathcal{J}^T}
e^{\frac{\Delta \tau}{2} \mathcal{D}^T} C(x,v,t).
\end{equation}
The fractional steps representation of this scheme is
\begin{align} \label{fwSplitFin}
C^{(1)}(x,v,t) &= e^{\frac{\Delta t}{2} \mathcal{D}^T } C(x,v,t), \\
C^{(2)}(x,v,t) &= e^{\Delta t \mathcal{J}^T} C^{(1)}(x,v,t) \nonumber, \\
C(x,v,t+\Delta t) &= e^{\frac{\Delta t}{2} \mathcal{D}^T } C^{(2)}(x,v,t).  \nonumber
\end{align}

\section{Backward and Forward Finite-Difference Scheme for the Diffusion Part} \label{fdscheme_diffusion}
\paragraph{Backward scheme.}
We follow \cite{HoutWelfert2007}, who consider the unconditional stability of second-order finite-difference schemes used to numerically solve multi-dimensional diffusion problems containing mixed spatial derivatives. They investigate the ADI scheme proposed by Craig and Sneyd (see references in the paper), a modified version of Craig and Sneyd's ADI scheme, and the ADI scheme introduced by Hundsdorfer and Verwer. Both necessary and sufficient conditions are derived on the parameters of each of these schemes for unconditional stability in the presence of mixed derivative terms. The main result of \cite{HoutWelfert2007} is that, under a mild condition on the parameter $\theta$ of the scheme, the second-order Hundsdorfer and Verwer (HV) scheme is unconditionally stable when applied to semi-discretized diffusion problems with mixed derivative terms in an arbitrary spatial dimension $k > 2$.

Following \cite{HoutWelfert2007}, consider the initial-boundary value problems
for two-dimensional diffusion equations, which after the spatial discretization lead to
initial value problems for huge systems of ordinary differential equations
\begin{equation}\label{initBoundProb}
    V'(\tau) = F(\tau,V(\tau)) \quad \tau \ge 0, \quad V(0) = V_0,
\end{equation}
\ni with given vector-valued function $F$ and initial vector $V_0$. \cite{HoutWelfert2007}
consider splitting schemes  for the numerical solution of the \eqref{initBoundProb}.
They assume that $F$ is decomposed into the sum
\begin{equation}\label{spl1}
F(\tau,V) = F_0(\tau, V) +F_1(\tau, V) + \cdots + F_k(\tau, V),
\end{equation}
where the $k+1$ terms $\{F_j\}$ are easier to handle than $F$ itself. The term $F_0$ contains all contributions to $F$ stemming from the mixed
derivatives in the diffusion equation, and this term is always treated explicitly in the numerical time integration. Next, for each $j \ge 1$, $F_j$ represents the contribution to
$F$ stemming from the second-order derivative in the $j$th spatial direction, and this term is always treated implicitly.

Further, \cite{HoutWelfert2007} analyze two splitting schemes, and one of them is the HV scheme. This scheme defines an approximation $V_n \approx V(\tau_n)$, $n=1,2,3, \dots$, by performing a series of (fractional) steps:
\begin{align} \label{HV}
Y_0 &= V_{n-1} + \Delta \tau F(\tau_{n-1},V_{n-1}), \\
Y_j &=  Y_{j-1} + \theta \Delta \tau \left[F_j(\tau_n,Y_j) - F_j(\tau_{n-1},V_{n-1})\right], \ j = 1,2,\dots,k, \nn \\
\tilde{Y}_0 &= Y_0 + \dfrac{1}{2} \Delta \tau \left[F(\tau_n,Y_k) - F(\tau_{n-1},V_{n-1})\right], \nn \\
\tilde{Y}_j &= \tilde{Y}_{j-1} + \theta \Delta \tau \left[F_j(\tau_n,\tilde{Y}_j) - F_j(\tau_n,Y_k)\right], \ j=1,2,\dots,k, \nn \\
V_n &= \tilde{Y}_k. \nn
\end{align}
This scheme is of order two in time  for any value of $\theta$, so this parameter can be chosen to meet additional requirements.

\cite{HoutWelfert2007} also investigate the stability of this scheme using von Neumann analysis. Accordingly, stability is always considered in the $L_2$ norm, and, in order to make the analysis feasible, all coefficients in \eqref{HV} are assumed to be constant and the boundary condition to be periodic. Under these assumptions, the matrices $A_0,A_1, \dots , A_k$ that are obtained by finite-difference discretization of operators $F_k$  are constant and form Kronecker products of circulant matrices.\footnote{An explicit discretization of $F_k$ in our case is discussed below.} Hence, the $\{A_k\}$ are normal and commute with each other. This implies that stability can be analyzed by considering the linear scalar ordinary differential equation
\begin{equation} \label{stab1}
V'(\tau) = (\lambda_0 + \lambda_1 + ... + \lambda_k)V(\tau),
\end{equation}
\ni where $\lambda_j$ denotes an eigenvalue of the matrix $A_j$,  $0 \le j \le k$. Then, by analyzing \eqref{stab1},  \cite{HoutWelfert2007} prove some important theorems to show
unconditional stability of their splitting scheme when $\theta \ge 1/3$.

An important property of this scheme is that the mixed derivative term in the first
equation of \eqref{HV} is treated explicitly while all further implicit steps
contain only derivatives in time and one spacial coordinate. In other words,
the entire 2D unsteady problem is reduced to a set of four 1D unsteady equations and two explicit equations.

For the semi-discretization of \eqref{HV}, the authors consider finite
differences. All spatial derivatives are approximated using second-order central
differences on a rectangular grid with constant mesh width $\Delta x_i
> 0$ in the $x_i$-direction ($1 \le i \le k$). Details of this scheme
implementation are discussed in \cite{HoutFoulon2010}. Their experiments
show that a choice of $\theta = 1/3$ for the Heston model  is good. They also demonstrate
that this scheme has a stiff order of convergence in time equal to 2.\footnote{In other words, the order of convergence does not fluctuate significantly with time step change, and is always very close to 2.}

It can easily be observed that the first and third equations in \eqref{oper} are of the same (Heston-like) type. Therefore, we will use the scheme described above when solving these equations.

\paragraph{Forward scheme.}
As discussed in Section \ref{introduction}, a consistent finite-difference scheme for the forward equation could be constructed by finding an explicit representation for the transposed discrete operator $\mathcal{A} = \mathcal{B}^T$ in \eqref{BKE1}. In our case, an explicit form of $\mathcal{B}$ is determined implicitly via \eqref{HV} because the latter admits the representation
\begin{equation} \label{R}
V(x, v, \tau + \Delta \tau) = R(\tau) V(x, v, \tau),
 \end{equation}
where $R$ is the corresponding transition matrix. To connect these two forms, let us formally integrate \eqref{BKE1} to obtain the solution represented via a matrix exponential:
\[ V(x, v, \tau + \Delta \tau) = e^{\Delta \tau \mathcal{B}} V(x, v, \tau)  = R V(x, v, \tau). \]
Thus, $R = e^{\Delta \tau \mathcal{B}}$. On the other hand, as $\exp(\Delta \tau \mathcal{B}^T) = (e^{\Delta \tau \mathcal{B}})^T$, the forward scheme could be represented
\begin{equation} \label{RT}
V(x, v, t + \Delta t) = R^T(t) V(x, v, t).
\end{equation}
Therefore, all we need to build a consistent forward scheme given the backward scheme \eqref{HV} is to construct an explicit form of $R^T$. As shown in the appendix, the result is
\begin{align} \label{BKdiscrOper}
R^T &= \left\{M_1^{-1}\left[I + \frac{1}{2} \Delta t \left(F^{n-1} + F^n R_2\right) - \theta \Delta t F^n_1 R_2\right] - \theta \Delta t F^n_2 R_2 \right\}^T (M_2^T)^{-1} \\
&= \left\{\left[I + \frac{1}{2} \Delta t \left((F^{n-1})^T + R^T_2 (F^n)^T \right) - \theta \Delta t R^T_2 (F^n_1)^T \right](M_1^T)^{-1} - \theta \Delta t R_2^T (F^n_2)^T \right\} (M_2^T)^{-1}, \nn \\
R_2^T &= \left\{\left[I + \Delta t \left((F^{n-1})^T - \theta (F^{n-1}_1)^T  \right) \right](M_1^T)^{-1} - \theta \Delta t (F_2^{n-1})^T \right\}(M_2^T)^{-1}, \nn
\end{align}
\noindent where we set $\Delta \tau = \Delta t$,  $M_i \equiv I - \theta\Delta t F^n_i$, and $F_i^n = F_i(\tau_n)$.

The expression in the right-hand side of \eqref{BKdiscrOper} could be simplified if one wants to derive a forward scheme consistent with the backward scheme up to a necessary order of approximation in $\Delta \tau$. However, we do not follow this approach in the present paper because, as we stated in the introduction, our goal here is slightly different, namely to derive a forward scheme that exactly matches its backward counterpart.

Based on this representation of $R^T$, the forward scheme now can be converted to a set of fractional steps, as occurs in \eqref{HV}. The algorithm is:
\begin{align} \label{fwAlgo}
M^T_2 Y_0 &= V_{n-1}, \\
M^T_1 Y_1 &= Y_0, \nn \\
\tilde{Y_0} &= \left[\frac{1}{2}(F^n)^T - \theta (F_1^n)^T\right] Y_1 - \theta (F_2^n)^T Y_0, \nn \\
M^T_2 \tilde{Y}_1 &= \tilde{Y}_0, \nn \\
M^T_1 \tilde{Y}_2 &= \tilde{Y}_1, \nn \\
V_n &= Y_1 + \Delta t \left\{
\frac{1}{2} (F^{n-1})^T Y_1 + \tilde{Y}_2 + \Delta t \Big[ \left[(F^{n-1})^T - \theta (F^{n-1}_1)^T \right]\tilde{Y}_2 - \theta (F_2^{n-1})^T \tilde{Y}_1\Big] \right\}. \nn
\end{align}

This scheme, however, has two problems. First, when using splitting (or fractional steps), one usually wants all internal vectors $Y_j$, $j=0,1$ and $\tilde{Y}_k$, $k=0,1,2$ to form consistent approximations to $V_n$. The scheme in \eqref{fwAlgo} loses this property at step 3. Second, because at step 3 the norm of the matrix on the right-hand side is small, the solution is sensitive to round-off errors.

To resolve these issues, one can do a trick---namely, represent \eqref{BKdiscrOper} in a form
\begin{align} \label{BKtrans}
V_n &= Y_1 + c^{n-1}_0 \Delta t Y_1  + \Delta t R^T_2 \left[ c^n_1 Y_1 - c^n_2 Y_0 \right] \\
c^n_0 &= \frac{1}{2} (F^{n})^T, \quad c^n_1 = \frac{1}{2}(F^n)^T - \theta (F_1^n)^T, \quad
c^n_2 = \theta (F_2^n)^T. \nonumber
\end{align}
Now add and subtract $R_2^T V_{n-1}$ to the right-hand side of \eqref{BKtrans}, and take into account that
\[ R_2^T V_{n-1} = Y_1 + \Delta t \left[c_3^{n-1} Y_1 - c^{n-1}_2 Y_0 \right], \qquad c_3^n = (F^{n})^T - \theta (F_1^n)^T. \]
Therefore, \eqref{fwAlgo} can now be written in the form
\begin{align} \label{fwAlgo_alt}
M^T_2 Y_0 &= V_{n-1}, \\
M^T_1 Y_1 &= Y_0, \nn \\
\tilde{Y_0} &= V_{n-1} + \Delta t \left(c_1^n Y_1 - c_2^n Y_0\right), \nn \\
M^T_2 \tilde{Y}_1 &= \tilde{Y}_0, \nn \\
M^T_1 \tilde{Y}_2 &= \tilde{Y}_1, \nn \\
V_n &= \tilde{Y}_2 + \Delta t \left[c_3^{n-1} \Delta \tilde{Y}_2 - c_2^{n-1} \Delta \tilde{Y}_1 + c_0^{n-1} Y_1\right], \nn
\end{align}
\noindent where $\Delta \tilde{Y}_i = \tilde{Y}_i - Y_{i-1}$, $i=1,2$.

Proceeding in a similar way, we can derive a forward analog for another popular backward finite-difference scheme---a modified Craig-Sneyd (MCS) scheme, see \cite{HoutFoulon2010}.  The backward and forward schemes are:
\paragraph{Backward scheme:}
\begin{align} \label{MCS}
Y_0 &= V_{n-1} + \Delta \tau F(\tau_{n-1},V_{n-1}), \\
Y_j &=  Y_{j-1} + \theta \Delta \tau \left[F_j(\tau_n,Y_j) - F_j(\tau_{n-1},V_{n-1})\right], \ j = 1,2,...,k \nn \\
\tilde{Y}_0 &= Y_0 + \frac{1}{2} \Delta \tau \left[F(\tau_n,Y_k) - F(\tau_{n-1},V_{n-1})\right] \nn \\
& - \theta \Delta \tau \left[F_1(\tau_n,Y_k) - F_1(\tau_{n-1},V_{n-1}) +
F_2(\tau_n,Y_k) - F_2(\tau_{n-1},V_{n-1}) \right], \nn \\
\tilde{Y}_j &= \tilde{Y}_{j-1} + \theta \Delta \tau \left[F_j(\tau_n,\tilde{Y}_j) - F_j(\tau_n,V_{n-1})\right], \ j=1,2,...,k \nn \\
V_n &= \tilde{Y}_k. \nn
\end{align}

\paragraph{Forward scheme:}
\begin{align} \label{fwMCS}
M^T_2 Y_0 &= V_{n-1}, \\
M^T_1 Y_1 &= Y_0, \nn \\
\tilde{Y_0} &= V_{n-1} + \Delta t c_-^n Y_1, \nn \\
M^T_2 \tilde{Y}_1 &= \tilde{Y}_0, \nn \\
M^T_1 \tilde{Y}_2 &= \tilde{Y}_1, \nn \\
V_n &= \tilde{Y}_2 + \Delta t \left\{
c_3^{n-1} \Delta \tilde{Y}_2 - c_2^{n-1} \Delta \tilde{Y}_1 + \left[c_+^{n-1} - \theta (F_1^n)^T\right]Y_1 - c_2^n Y_0\right\}, \nn
\end{align}
\noindent where
\[ c_+^{n} = \frac{1}{2} (F^n)^T + \theta \left[(F_1^n)^T + (F_2^n)^T \right], \quad c_-^n = \frac{1}{2} (F^n)^T - \theta \left[(F_1^n)^T + (F_2^n)^T \right]. \]

The boundary conditions for the forward scheme should be consistent with those for the backward scheme.\footnote{Here for simplicity we consider only Dirichlet boundary conditions.} However, these two sets of conditions will not be exactly the same. Indeed, in the forward equation the dependent variable is the density, while in the backward equation the dependent variable is the undiscounted option price. For the latter, boundary conditions are set to the payoff function, while the boundary conditions are set to the density function for the former. Therefore, these boundary conditions could be quite different. For instance, for the put European option at $S=0$, we can set $V = K$, while for the density function this is $V=0$. Also, setting the boundary conditions in the $v$-domain should be done with a care. See \cite{Lucic2008} for a discussion on this matter.

\section{Forward Scheme for Jumps} \label{forwardscheme_jumps}
Efficient finite-difference schemes needed to discretize the operator $\mathcal{J}$ in \eqref{splitFin} (for the backward equation) on a given grid were provided in \cite{Itkin2013} for some popular jump models, such as Merton, Kou, and CGMY. The proposed method is almost universal (i.e., it allows computation of PIDEs for various jump-diffusion models in a unified form), while implementation of the method is relatively simple. It was shown that for the Merton and Kou models, the matrix exponential could be computed up to second-order in the space-grid step size $h$ using P{\'a}de approximations and Picard fixed-point iterations, with the total complexity $O(N)$, where $N$ is the number of space nodes. It was also proved that
these schemes (i) are unconditionally stable, (ii) provide second-order in $h$,\footnote{The first order schemes were also considered in the paper} and (iii) preserve the positivity of the solution. The results are presented as propositions, and the corresponding proofs are given based on modern matrix analysis, including a theory of M-matrices, Metzler matrices, and eventually exponentially positive matrices.

A slight modification of the method makes it possible to apply it to the CGMY model with a parameter $\alpha > 0$.\footnote{For $\alpha \le 0$, this class of schemes was proposed in \cite{ItkinCarr2012Kinky}.} This becomes especially important for $\alpha \ge 1$ because in this range, known similar algorithms (e.g., \cite{WangWanForsyth2007}) experience problems. Analysis made in \cite{Itkin2013} demonstrates a convergence close to $O(h^{1.8})$ numerically, despite the theoretical convergence of $O(h^2)$. One possible reason for this discrepancy could be that the maximum eigenvalue of the transition matrix in this case is close to 1, and therefore the roundoff errors could be important. Performance-wise, computing the matrix exponential in the matrix form followed by direct multiplication of this matrix by a vector of solutions seems to be more efficient than the FFT-based algorithm in \cite{WangWanForsyth2007}, as Picard iterations in this case converge pretty slowly.

Using this new method, there is no issue with computing $\mathcal{J}^T$. Indeed, if $\mathcal{J}$ is the negative of an M-matrix, the transposed matrix preserves the same property. Also if $\mathcal{J}$ has all negative eigenvalues, the same is true for $\mathcal{J}^T$. Then the unconditional stability of the scheme and its property of preserving the positivity of the solution follow from Proposition 4.1 in \cite{Itkin2013}.

\section{Implementation Details and Numerical Experiments} \label{numerical_results}
In this section, we discuss some points that could be important when implementing the above approach.

\subsection{Damping}
It is known that the temporal error at the first time steps could be relatively large, especially for modest time steps; see \cite{Rannacher:1984, HoutFoulon2010} and references therein. This is caused by the fact that for the backward scheme at $S = K$,  the payoff function (say, for a European call or put) is not a smooth function of $S$, and many finite-difference schemes do not sufficiently dampen local high-frequency errors. To resolve this, it was proposed in \cite{Rannacher:1984} to make the first few temporal steps with fully implicit Euler scheme (which is of the first order approximation in time) and then switching to a scheme of your choice.

For the corresponding forward equation, this situation is even more pronounced, since at $t=0$ the initial density is a Dirac delta function $\delta(S-S_0)\delta(v-v_0)$. Therefore, it is desirable to apply damping to the first few steps of the forward induction.

Thus, for consistency the damping algorithm should be used at both ends of the time grid, for instance at first two steps $t_1, t_2$ and last two steps $t_{M-1}, t_M$ of the forward induction, and at first two steps $\tau_1, \tau_2$ and last two steps $\tau_{M-1}, \tau_M$ of the backward induction. Accordingly, the whole implicit Euler operator (matrix) for the forward scheme should be a transposed operator (matrix) of that for the backward scheme.

\subsection{Nonuniform Grids}
Nonuniform grids are used to increase computational accuracy of pricing. A standard choice is to make the grid finer at regions where the solution gradients are high. For example, when pricing European options, it makes sense to have a fine grid close to the strike of the deal. Therefore, this argument could be used for the backward algorithm, which amounts to pricing.

For the forward approach, used mainly for calibration, intuition tells us to use a finer grid close to the initial point $(S_0, v_0)$. Thus, the choice of regions where the grid should condense differs in nature for the backward and forward equations. Moreover, after the transition density is found by solving the forward equation, it can be used to simultaneously get prices for all options with same maturity $T$ and $S_0$ and different strikes $K$. However, pricing of these options requires multiple runs of the backward algorithm.

Therefore, to have all these arguments consistent with each other, one possible solution is to produce a grid that is fine near all the needed strikes\footnote{Assuming that we know all necessary strikes in advance.} and the initial point. In other words, this grid will have multiple areas with fine space steps alternated with areas with coarse space steps.

Technically, this could be achieved by using a grid construction algorithm proposed in \cite{fdm2000}.

Also for the forward grid, we need the initial point $(S_0,v_0)$ to be on the grid for a better representation of the initial condition. However, for the backward equation, this is not necessary. Moreover, it is recommended to have a strike in the middle of the corresponding grid cell, not at the grid node, for better accuracy.  In our case, this could cause a problem when constructing a grid that condenses at multiple strikes. This problem, however, could be eliminated following the idea of \cite{fdm2000}; see also \cite{Tinne2013}. Given a strike, we replace the value of the payoff function at the grid point $S_i$ nearest to the strike $K$ with its average over the cell centered at $S_i$:
\[\mathit{Payoff}(S_i,K) =   \frac{1}{h} \int_{S_{i-1/2}}^{S_{i+1/2}} \mbox{Payoff}(s,K)ds, \]
\noindent where $S_{i - 1/2} = \frac{1}{2}(S_{i-1} + S_i), \ S_{i + 1/2} = \frac{1}{2}(S_{i} + S_{i+1}), \ h = S_{i+ 1/2} - S_{i - 1/2}$. This allows an efficient reduction of the discretization error close to the strike, while eliminating unnecessary complications in building the grid.

\subsection{Positivity of the Solution}
Because the solution is expected to be a valid approximation of $V_n$ at every step of the splitting scheme, all the vectors $Y_j$, $j=0,1$, and $\tilde{Y}_k$, $k=0,1,2$, should be nonnegative. That means that at every fractional step of the scheme, the corresponding operators must preserve the positivity of the solution.

For steps 1, 2, and 4 in \eqref{fwAlgo}, this is guaranteed if both $M_1^T$ and  $M_2^T$ are M-matrices; see \cite{BermanPlemmons1994}. To achieve this, an appropriate (upward) approximation of the drift (convection) term has to be chosen, which is often discussed in the literature; see \cite{HoutFoulon2010} and references therein.

For steps 3 and 5, this is a more delicate issue. A seven-point stencil for discretization of the mixed-derivative operator that preserves the positivity of the solution was proposed in \cite{toivanen2010, chiarella2008} for correlations $\rho < 0$, and in \cite{IkonenToivanen2008, IkonenToivanen2007} for positive correlations. However, in their schemes the mixed derivative term was treated implicitly (that is the reason they needed a discretized matrix to be an M-matrix). In our case the entire matrix in the right-hand side of steps 3,5 should be either a positive matrix, or a Metzler matrix (in this case the negative of an M-matrix). The latter can be achieved when using approximations of \cite{toivanen2010, chiarella2008} and \cite{IkonenToivanen2008, IkonenToivanen2007} in an opposite order, i.e. use approximations recommended for $\rho > 0$ when $\rho < 0$, and vice versa.

This approach, however, puts some constraints on the grid steps $h_i$, $i=1,\dots, N$. Therefore, such discretization should be considered together with the algorithm of building a nonuniform grid. It is usually better to use a uniform grid in $v$-space, and then it is easier to obey the positivity constraints on the steps in $S$-space.

Also, our experiments showed that the forward scheme is much more sensitive to the choice of discretization of the mixed derivative than the backward one.

\subsection{Discrete Dividends}
\begin{figure}[h!]
\begin{minipage}{0.46\linewidth}
\begin{center}
\fbox{\includegraphics[width=\linewidth]{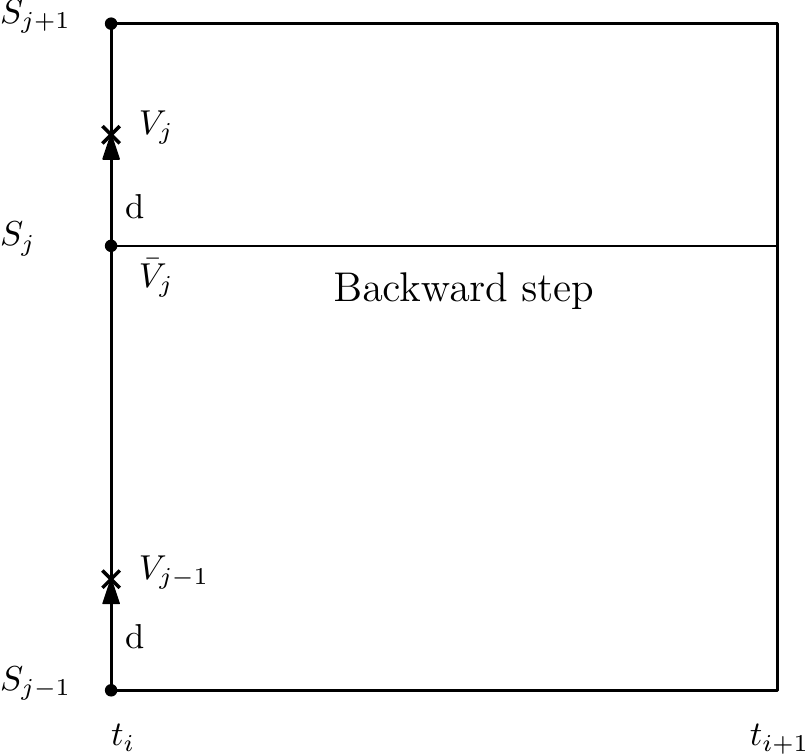}}
\caption{Grid shift to account for discrete dividends in a backward approach.}
\label{bkDivs}
\end{center}
\end{minipage}
\hspace{0.04\linewidth}
\begin{minipage}{0.46\linewidth}
\begin{center}
\fbox{\includegraphics[width=\linewidth]{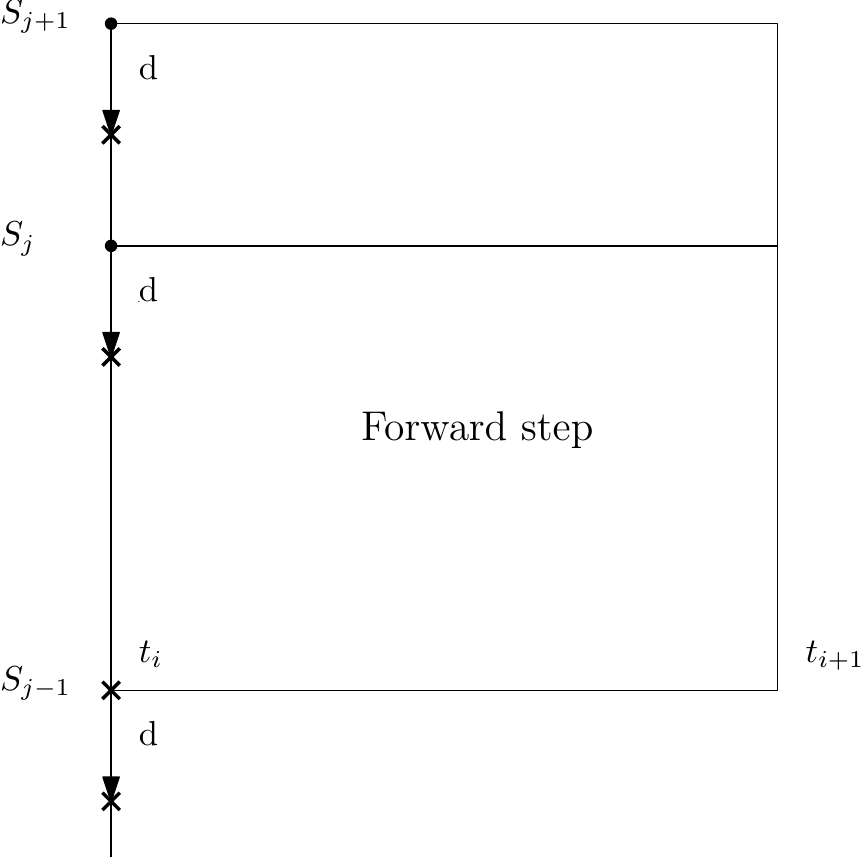}}
\caption{Grid shift to account for discrete dividends in a forward approach.}
\label{fwDivs}
\end{center}
\end{minipage}
\end{figure}

A standard way to account for discrete dividends when using tree or lattice methods is a shift of grid (space states) at an ex-dividend date; see, e.g., \cite{hull:97}.   This approach is illustrated in Fig.~\ref{bkDivs} for the backward scheme and in Fig.~\ref{fwDivs} for the forward approach.

A more detailed description of the algorithm follows.

Suppose we solve a backward PDE. At time step $t_i$, which corresponds to an ex-dividend date,\footnote{If the ex-dividend date falls between two nodes of the temporal grid, a common approach is to move it to the grid node either forward (and adjusting the dividend amount to the forward value of the dividend at that date) or backward (and discounting the dividend amount back to that date).} a new value of the option price $V_i$ is found by applying a backward finite-difference algorithm. Therefore, $V_i({\bf S}) = R V_{i-1}({\bf S})$. After that, since we go backward, the option values at time $\tau^+_i = \tau_i + \epsilon$, $\epsilon \ll 1$, which is just before the dividends were paid, become $V_{i}({\bf S}+d)$.  That is, the same option values now belong to the grid nodes shifted up by the dividend amount $d$. This relies on the fact that the option values are continuous at the moment when a dividend is paid. As it is not convenient to work with a shifted grid, usually practitioners reinterpolate the new option values back to the previous grid. So if we let  $\mathcal{I}_u$ denote  a corresponding interpolation operator with shifts up, the final option values at the moment $\tau^+_{i}$ are $\bar{V}_i = \mathcal{I}_b R V_{i-1}$.

For the forward equation, taking into account the way we constructed a consistent forward algorithm, this expression should be $\bar{V}_i = [\mathcal{I} R]^T V_{i-1} = R^T \mathcal{I}^T V_{i-1}$. This means that: i) when moving forward in time $t$ and giving the options values $V_{i-1}$ at time $t^i_i = t_i - \epsilon$, we need first to shift the grid ${\bf S}$ down by $d$, so that $V_{i-1}$ now belongs to the nodes of the shifted grid, $V_{i-1}(\mathbf{S}-d)$; ii) then we need to  reinterpolate them back to the original grid using a corresponding interpolation operator with shifts down $\mathcal{I}_d$, thus obtaining $\bar{V}_{i-1}({\bf S})$; and iii) we finally need to apply the operator $R^T$ to get the final option values $V_i({\bf S})$.

The main issue here is to construct interpolation operators such that $\mathcal{I}_u = \mathcal{I}^T_d$. For simplicity, assume first that the dividend amount $d$ is less than any step $h_i$, $i=1, \dots,N$, of the grid. As the  finite-difference schemes discussed above are of second-order approximation in $h$, linear interpolation is sufficient. Therefore, in the backward approach after the values $V_{i} = R V_{i-1}$ are obtained at time step $t_i$ and shifted to the grid ${\bf S}+d$ (see Fig.~\ref{bkDivs}), the interpolated value $\bar{V}_i(S_j)$ is
\begin{equation} \label{intBK}
\bar{V}_i(S_j) = V_{j-1} \frac{d}{h_{j,j-1}} + V_j \frac{h_{j,j-1} - d}{h_{j,j-1}}, \qquad h_{j,j-1} = S_j - S_{j-1}.
\end{equation}
Therefore, the matrix $B$ of the operator $\mathcal{I}_b$ discretized at this grid is
\[ B = \left(
     \begin{array}{ccccc}
       B_{00} & 0 & 0 & 0 & ... \\
       \frac{d}{h_{10}} & \frac{h_{10}-d}{h_{10}} & 0 & 0 & ... \\
       0 & \frac{d}{h_{21}} & \frac{h_{21}-d}{h_{21}} & 0 & ... \\
       ... & ... & ... & ... & ... \\
       ... & 0 & \frac{d}{h_{N-1,N-2}} & \frac{h_{N-1,N-2}-d}{h_{N-1,N-2}} & 0  \\
       ... & 0 & 0 & B_{N,N-1} & B_{N,N} \\
     \end{array}
   \right).
\]
This is a lower bi-diagonal matrix where the elements $B_{00}$ in the first row and $B_{N,N-1}$, $B_{N,N}$ in the last row are determined by the boundary conditions on $V_i(S)$ at $S_\mathrm{min}$ and $S_\mathrm{max}$. Proceeding in a similar way with the forward approach (see Fig.~\ref{fwDivs}), we find the corresponding matrix $F$ for the operator $\mathcal{I}_f$:
\[ F = \left(
     \begin{array}{ccccc}
       F_{00} & F_{01} & 0 & 0 & ... \\
       0 & \frac{h_{21}-d}{h_{21}} & \frac{d}{h_{21}} & 0 & ... \\
       ... & ... & ... & ... & ... \\
       ... & 0 & 0 &  \frac{h_{N-1,N-2}-d}{h_{N-1,N-2}} & \frac{d}{h_{N-1,N-2}}  \\
       ... & 0 & 0 & 0 & F_{N,N} \\
     \end{array}
   \right).
\]
This is an upper bi-diagonal matrix where the elements $F_{00}$, $F_{01}$ in the first row and $F_{N,N}$ in the last row are determined by the boundary conditions on $V_i(S)$ at $S_\mathrm{min}$ and $S_\mathrm{max}$.

As was explained above, we need $F = B^T$. Clearly, this is not the case unless the grid is uniform and $h_{i,i-1} = h_{i+1,i}$  for $i= 2, \dots, N-1$. But in many situations this is impractical. One way to eliminate this difficulty if one wants to work with a nonuniform grid is to eliminate re-interpolation, and after the dividend is paid, proceed to work with a shifted grid. This, however, also brings some technical problems with setting boundary conditions.

Another problem with the forward approach is that $S-d$ becomes negative when $S$ is small. An extended discussion of this situation can be found in \cite{HHLewis2003}. The main idea is that if $S < d$ the company cannot pay a dividend $d$, but at most the amount $S$. Therefore, in this case we assume the dividend $D$ is now a function of $S$, i.e., $D(S) = d$, when $S > d$, and $D(S) = S$ when $0 \le S \le d$.

\subsection{Parameters of the Finite-Difference Scheme}
Both MCS and HV backward finite-difference schemes are investigated in literature.  In \cite{HoutWelfert2007} and \cite{HoutFoulon2010}, the authors apply the methods to the Heston model for convergence and stability. It is shown that the MCS scheme with parameter $\theta = 1/3$ and damping at $\tau = 0$ provides a fast, accurate, and robust numerical solution with arbitrary correlation $\rho\in [-1,1]$. The original Craig-Sneyd scheme (see \cite{CraigSneyd1988}) with $\theta = 1/2$ and the HV scheme with $\theta = 1/2 + \sqrt{3}/6$ with damping
at $\tau=0$ also provide a good alternative. All three splitting schemes demonstrate an unconditional stability and a stiff order of convergence equal to two.

Here we want to investigate how the choice of $\theta$ in the forward scheme affects the convergence and stability of the solution. We consider a call option and take the Heston model parameters as shown in Table~\ref{TabTest}.
\begin{table}[h!]
\begin{center}
\begin{tabular}{|c|c|c|c|c|c|c|c|c|c|c|c|c|}
\hline
$T$ & $K$ & $r$ & $q$ & $\xi$ & $\rho$ & $\kappa$ & $v_{\infty}$   \cr
\hline
1.0 & 100 & 0.05 & 0.0 & 0.3 & 0.8 & 1.5 & 0.1    \cr
 \hline
\end{tabular}
\end{center}
\caption{Initial parameters used in test calculations.}
\label{TabTest}
\end{table}
Here $r$ is the interest rate and $q$ is the dividend yield. As we considered the Heston model, the local volatility function was set to $\phi(S_t,t) = 1$.

A nonuniform space grid was constructed in the $S$-dimension with 76 nodes in $S \in [0,S_\mathrm{max}]$, $S_\mathrm{max} = 40 \max(S_0,K)$, and a uniform grid was constructed with 79 nodes in $v \in [0, v_\mathrm{max}]$, $v_\mathrm{max} = 6 v_0$. Here $K$ is the strike. A further increase of $S_\mathrm{up}$ does not influence the option price much, so this boundary was chosen based on a practical argument. We also used 100 time steps.

We solved both the forward and the backward equations and compared the prices obtained as a function of the parameter $\theta$. The results for the HV scheme are presented in Table~\ref{Tab1}. The benchmark (analytical) solution $C_\mathrm{FFT}$ was obtained using the FFT; see \cite{CarrMadan:99a}. The relative error $\epsilon_\mathrm{bk}$ obtained by pricing this option using a backward approach, and $\epsilon_\mathrm{fw}$ obtained by using the forward approach are determined as $\epsilon_i = (C_\mathrm{FFT} - C_i)/C_\mathrm{FFT} \cdot 100\%$ for $i \in \{\mathrm{bk},\mathrm{fw}\}$. In this test, we used $S_0 = 100$, $v_0 = 0.5$, which gives $C_\mathrm{FFT} = 24.0047$.
\begin{table}[h!]
\begin{center}
\begin{tabular}{|c|r|r|r|r|r|r|r|r|r|r|r|r|}
\hline
$\theta$  & 0.3 & 0.4 & 0.5 & 0.6 & 0.7 & 0.8 & 0.9 & 1.0
\cr
\hline
$\epsilon_\mathrm{bk}$ & 0.0718 & 0.0719 & 0.0720 & 0.0719 & 0.0717 & 0.0714 & 0.0709 & 0.0704
\cr
\hline
$\epsilon_\mathrm{fw}$ & 0.0595 & 0.0597 & 0.0597 & 0.0597 & 0.0595 & 0.0592 & 0.0588 & 0.0583
\cr
\hline
$\epsilon_\mathrm{bk} - \epsilon_\mathrm{fw}$ & 0.0123 &  0.0122 &  0.0123 &  0.0122 &  0.0122 &  0.0122 &  0.0121 &  0.0121
\cr
\hline
\end{tabular}
\end{center}
\caption{Convergence of the HV backward and forward solutions as a function of $\theta$.}
\label{Tab1}
\end{table}
The option prices from the forward and backward methods agree up to 1 bp. The value of $\theta$ where $C_\mathrm{bk}$ and $C_\mathrm{fw}$ provide the best approximation to $C_\mathrm{FFT}$ among all trials of $\theta$ is 1. The latter, however, contradicts conclusions drawn in \cite{HoutWelfert2007} and \cite{HoutFoulon2010}, where the value $\theta = 1/2 + \sqrt{6}/3$ is recommended.

Similar results for the MCS scheme are given in Table~\ref{Tab3}. Here again both backward and forward schemes demonstrate a better approximation to the FFT value at $\theta = 1$, while in \cite{HoutFoulon2010} it is reported that the value $\theta = 1/3$ is the most robust.

\begin{table}[h!]
\begin{center}
\begin{tabular}{|c|r|r|r|r|r|r|r|r|r|r|r|r|}
\hline
$\theta$  & 0.3 & 0.4 & 0.5 & 0.6 & 0.7 & 0.8 & 0.9 & 1.0
\cr
\hline
$\epsilon_\mathrm{bk}$ & 0.0718 & 0.0719 & 0.0720 & 0.0719 & 0.0717 & 0.0713 & 0.0709 & 0.0703
\cr
\hline
$\epsilon_\mathrm{fw}$ & 0.0595 & 0.0596 & 0.0597 & 0.0596 & 0.0594 & 0.0590 & 0.0586 & 0.0581
\cr
\hline
$\epsilon_\mathrm{bk} - \epsilon_\mathrm{fw}$ & 0.0123 &   0.0123 &   0.0123 &   0.0123 &   0.0123 &   0.0123 &   0.0123 &   0.0122
\cr
\hline
\end{tabular}
\end{center}
\caption{Convergence of the MCS backward and forward solutions as a function of $\theta$.}
\label{Tab3}
\end{table}

Based on this experiment, we can conclude that the convergence of the backward and forward schemes is similar if the same value of the parameter $\theta$ is chosen for both algorithms.

As was already mentioned, we use a special discretization of the mixed derivative operator $F_0$ to have the corresponding matrix be an M-matrix. This discretization is of the second-order of approximation in the grid steps in $S$ and $v$ directions when the point $(S_i, v_j)$ is far from the boundaries of the computational domain. However, at the boundaries the approximation drops down to the first order. Most likely this introduces some additional error that does not exist in the uncorrelated case, i.e., when $\rho = 0$. Therefore, to check this and also the convergence of both the forward and backward methods in this case, we run the same tests now with $\rho = 0$, where $C_\mathrm{FFT} =  23.7015$.  The results for the HV and MCS schemes are presented in Table~\ref{Tab2} and Table~\ref{Tab4}, respectively.
\begin{table}[h!]
\begin{center}
\begin{tabular}{|c|r|r|r|r|r|r|r|r|r|r|r|r|}
\hline
$\theta$  & 0.3 & 0.4 & 0.5 & 0.6 & 0.7 & 0.8 & 0.9 & 1.0
\cr
\hline
$\epsilon_\mathrm{bk}$ &  -0.0882 & -0.0879 & -0.0878 & -0.0878 & -0.0879 & -0.0881 & -0.0884 &
-0.0887
\cr
\hline
$\epsilon_\mathrm{fw}$ &  -0.0861 & -0.0861 & -0.0862 & -0.0863 & -0.0865 & -0.0868 & -0.0871 &
-0.0875
\cr
\hline
$\epsilon_\mathrm{bk} - \epsilon_\mathrm{fw}$ & -0.0021 &  -0.0018 &  -0.0016 &  -0.0015 &  -0.0014 &  -0.0013 &  -0.0013 &  -0.0012
\cr
\hline
\end{tabular}
\end{center}
\caption{Convergence of the HV backward and forward solutions as a function of $\theta$. Uncorrelated case.}
\label{Tab2}
\end{table}

\begin{table}[h!]
\begin{center}
\begin{tabular}{|c|r|r|r|r|r|r|r|r|r|r|r|r|}
\hline
$\theta$  & 0.3 & 0.4 & 0.5 & 0.6 & 0.7 & 0.8 & 0.9 & 1.0
\cr
\hline
$\epsilon_\mathrm{bk}$ &  -0.0888 & -0.0888 & -0.0889 & -0.0890 & -0.0893 & -0.0897 & -0.0902 &
-0.0907
\cr
\hline
$\epsilon_\mathrm{fw}$ &  -0.0861 & -0.0861 & -0.0862 & -0.0863 & -0.0865 & -0.0868 & -0.0872 &
-0.0876
\cr
\hline
$\epsilon_\mathrm{bk} - \epsilon_\mathrm{fw}$ & -0.0027 &  -0.0027 &  -0.0027 &  -0.0027 &  -0.0028   & -0.0029 &  -0.0030 &  -0.0031
\cr
\hline
\end{tabular}
\end{center}
\caption{Convergence of the MCS backward and forward solutions as a function of $\theta$. Uncorrelated case.}
\label{Tab4}
\end{table}

We see that forward and backward schemes produce similar results for all values of $\theta$, and the error with the FFT value is almost 4 times less (0.3 bp) then for the correlated case, which supports our analysis above. Here, however, the most robust value of $\theta$ for the backward HV scheme is 0.6 (which is close to that reported in
\cite{HoutWelfert2007}), and 0.3 for the forward scheme. For the backward MCS scheme, the most robust value is $\theta = 0.3$ ((which is also close to that reported in
\cite{HoutWelfert2007}).  The most robust value for the  forward  MCS scheme  is also $\theta = 0.3$.

One can also observe that in both cases ($\rho = 0$ and $\rho > 0$), the HV scheme provides slightly better results for both backward and forward schemes. The FFT option value is $C_{FFT} = 23.4077$.

Finally, the same experiments were repeated with $\rho = -0.8$. The results for the HV model are given in Tab.~\ref{Tab5}, and in Tab.~\ref{Tab6} for the MCS model.
\begin{table}[h!]
\begin{center}
\begin{tabular}{|c|r|r|r|r|r|r|r|r|r|r|r|r|}
\hline
$\theta$  & 0.3 & 0.4 & 0.5 & 0.6 & 0.7 & 0.8 & 0.9 & 1.0
\cr
\hline
$\epsilon_{bk}$ & -0.0801 & -0.0799 & -0.0798 & -0.0798 & -0.0798 & -0.0799 & -0.0801 & -0.0803
\cr
\hline
$\epsilon_{fw}$ & -0.0549 & -0.0545 & -0.0543 & -0.0541 & -0.0540 & -0.0541 & -0.0542 &
-0.0544
\cr
\hline
$\epsilon_{bk} - \epsilon_{fw}$ &  -0.0252 &  -0.0254 &  -0.0255 &  -0.0257 &  -0.0258 &  -0.0258 &  -0.0259 &  -0.0259
\cr
\hline
\end{tabular}
\end{center}
\caption{Convergence of the HV backward and forward solutions as a function of $\theta$ at $\rho = -0.8$.}
\label{Tab5}
\end{table}

\begin{table}[h!]
\begin{center}
\begin{tabular}{|c|r|r|r|r|r|r|r|r|r|r|r|r|}
\hline
$\theta$  & 0.3 & 0.4 & 0.5 & 0.6 & 0.7 & 0.8 & 0.9 & 1.0
\cr
\hline
$\epsilon_{bk}$ & -0.0801 & -0.0799 & -0.0798 & -0.0798 & -0.0799 & -0.0800 & -0.0802 & -0.0804
\cr
\hline
$\epsilon_{fw}$ & -0.0546 & -0.0542 & -0.0539 & -0.0538 & -0.0538 & -0.0540 & -0.0542 &
-0.0546
\cr
\hline
$\epsilon_{bk} - \epsilon_{fw}$ & -0.0255 & -0.0257 &  -0.0259  & -0.0260 &  -0.0261 &   -0.0260 &  -0.0260 &  -0.0258
\cr
\hline
\end{tabular}
\end{center}
\caption{Convergence of the MCS backward and forward solutions as a function of $\theta$ at $\rho = -0.8$.}
\label{Tab6}
\end{table}

Here, the HV and MCS schemes for both the forward and backward approaches are more accurate at $\theta$ around 0.7.

\clearpage
\section*{Acknowledgments}
I thank Peter Carr, Alex Lipton and Michael Konikov for some fruitful discussions.
I am indebted to Gregory Whitten, Steven O'Hanlon and Ben Meyvin for supporting this work.

\def\myBib{C:/NxData/AItkin/Papers/MySettings2011/aitkin_fin}
\bibliographystyle{authordate1}

\begin{thebibliography}{}

\bibitem[\protect\citename{Andersen \& Andreasen, }2000]{AA2000}
Andersen, L., \& Andreasen, J. 2000.
\newblock Jump diffusion processes: volatility smile fitting and numerical
  methods for option pricing.
\newblock {\em Review of Derivatives Research}, {\bf 4}, 231--262.

\bibitem[\protect\citename{Andreasen \& Huge, }2011]{AndreassenHuge2011}
Andreasen, J., \& Huge, B. 2011.
\newblock Random grids.
\newblock {\em Risk}, July, 66--71.

\bibitem[\protect\citename{Berman \& Plemmons, }1994]{BermanPlemmons1994}
Berman, A., \& Plemmons, R. 1994.
\newblock {\em Nonnegative matrices in mathematical sciences}.
\newblock SIAM.

\bibitem[\protect\citename{Carr \& Madan, }1999]{CarrMadan:99a}
Carr, Peter, \& Madan, Dilip. 1999.
\newblock Option Valuation Using the Fast Fourier Transform.
\newblock {\em Journal of Computational Finance}, {\bf 2}(4), 61--73.

\bibitem[\protect\citename{Chiarella {\em et~al.}, }2008]{chiarella2008}
Chiarella, C., Kang, B., Mayer, G.H., \& Ziogas, A. 2008.
\newblock {\em The Evaluation of American Option Prices Under Stochastic
  Volatility and Jump-Diffusion Dynamics Using the Method of Lines}.
\newblock Tech. rept. Research paper 219. Quantitative Finance Research Centre,
  University of Technology, Sydney.

\bibitem[\protect\citename{Cont \& Tankov, }2004]{ContTankov}
Cont, R., \& Tankov, P. 2004.
\newblock {\em Financial modelling with jump processes}.
\newblock Financial Matematics Series, Chapman \& Hall /CRCl.

\bibitem[\protect\citename{Craig \& Sneyd, }1988]{CraigSneyd1988}
Craig, I. J.~D., \& Sneyd, A.~D. 1988.
\newblock An alternating-direction implicit scheme for parabolic equations with
  mixed derivatives.
\newblock {\em Comp. Math. Appl.}, {\bf 16}, 341--350.

\bibitem[\protect\citename{de~Lange \& Raab, }1992]{OMQM}
de~Lange, O.~L., \& Raab, R.~E. 1992.
\newblock {\em Operator Methods in Quantum Mechanics}.
\newblock Oxford science publications.
\newblock Chapter 3.

\bibitem[\protect\citename{Eberlein, }2009]{Eberlein2009}
Eberlein, E. 2009.
\newblock Jump-type {L\'e}vy processes.
\newblock {\em Pages  439--455 of:} Andersen, T.~G., Davis, R.~A., Krei\ss,
  J.-P., \& Mikosch, T. (eds), {\em Handbook of Financial Time Series}.
\newblock Springer Verlag.

\bibitem[\protect\citename{Glasserman, }2003]{PGbook}
Glasserman, P. 2003.
\newblock {\em Monte Carlo Methods in Financial Engineering}.
\newblock Stochastic Modelling and Applied Probability, vol. 53.
\newblock Springer.

\bibitem[\protect\citename{Goodman, }2004]{Goodman2004}
Goodman, J. 2004.
\newblock {\em Forward and Backward Equations for Markov chains}.
\newblock available at
  \url{http://www.math.nyu.edu/faculty/goodman/teaching/StochCalc2004/notes/stoch_2.pdf}.

\bibitem[\protect\citename{Haentjens, }2013]{Tinne2013}
Haentjens, T. 2013.
\newblock Efficient and stable numerical solution of the
  Heston-–Cox-–Ingersoll-–Ross partial differential equation by
  alternating direction implicit finite difference schemes.
\newblock {\em International Journal of Computer Mathematics}, {\bf 90}(11),
  2409--2430.

\bibitem[\protect\citename{Haug {\em et~al.}, }2003]{HHLewis2003}
Haug, E., Haug, J., \& Lewis, A. 2003.
\newblock Back to basics: a new approach to the discrete dividend problem.
\newblock {\em Wilmott magazine}, {\bf September}, 37--47.

\bibitem[\protect\citename{Hout \& Foulon, }2010]{HoutFoulon2010}
Hout, K. J.~{In' t}, \& Foulon, S. 2010.
\newblock {ADI} finite difference schemes for option pricing in the {H}eston
  model with correlation.
\newblock {\em International journal of numerical analysis and modeling}, {\bf
  7}(2), 303--320.

\bibitem[\protect\citename{Hull, }1997]{hull:97}
Hull, John~C. 1997.
\newblock {\em Options, Futures, and other Derivative Securities}. third edn.
\newblock Upper Saddle River, NJ: Prentice-Hall, Inc.

\bibitem[\protect\citename{Ikonen \& Toivanen, }2007]{IkonenToivanen2007}
Ikonen, S., \& Toivanen, J. 2007.
\newblock Componentwise splitting methods for pricing American options under
  stochastic volatility.
\newblock {\em Int. J. Theor. Appl. Finance}, {\bf 10}, 331--361.

\bibitem[\protect\citename{Ikonen \& Toivanen, }2008]{IkonenToivanen2008}
Ikonen, S., \& Toivanen, J. 2008.
\newblock Efficient numerical methods for pricing American options under
  stochastic volatility.
\newblock {\em Num. Meth. PDEs}, {\bf 24}, 104--126.

\bibitem[\protect\citename{{In't Hout} \& Welfert, }2007]{HoutWelfert2007}
{In't Hout}, K.~J., \& Welfert, B.~D. 2007.
\newblock Stability of {ADI} schemes applied to convection-–diffusion
  equations with mixed derivative terms.
\newblock {\em Applied Numerical Mathematics}, {\bf 57}, 19--35.

\bibitem[\protect\citename{Itkin, }2013]{Itkin2013}
Itkin, A. 2013.
\newblock {\em Efficient Solution of Backward Jump-Diffusion PIDEs with
  Splitting and Matrix Exponentials}.
\newblock available at \url{http://arxiv.org/abs/1304.3159}.

\bibitem[\protect\citename{Itkin \& Carr, }2012]{ItkinCarr2012Kinky}
Itkin, A., \& Carr, P. 2012.
\newblock Using pseudo-parabolic and fractional equations for option pricing in
  jump diffusion models.
\newblock {\em Computational Economics}, {\bf 40}(1), 63--104.

\bibitem[\protect\citename{Koch \& Thalhammer, }2011]{ThalhammerKoch2010}
Koch, O., \& Thalhammer, M. 2011.
\newblock {\em Embedded Exponential Operator Splitting Methods for the Time
  Integration of Nonlinear Evolution Equations}.
\newblock Tech. rept. Institute for Analysis and Scientific Computing, Vienna
  University of Technology.

\bibitem[\protect\citename{Lanser \& Verwer, }1999]{LanserVerwer}
Lanser, D., \& Verwer, J.G. 1999.
\newblock Analysis of operator splitting for advection-diffusion-reaction
  problems from air pollution modelling.
\newblock {\em Journal of Computational and Applied Mathematics}, {\bf
  111}(1-2), 201--216.

\bibitem[\protect\citename{Lewis, }2000]{Lewis:2000}
Lewis, Alan~L. 2000.
\newblock {\em Option Valuation under Stochastic Volatility}.
\newblock Newport Beach, California, USA: Finance Press.

\bibitem[\protect\citename{Lipton, }2001]{Lipton2001}
Lipton, A. 2001.
\newblock {\em Mathematical Methods For Foreign Exchange: A Financial
  Engineer's Approach}.
\newblock World Scientific.

\bibitem[\protect\citename{Lipton, }2002]{Lipton2002}
Lipton, A. 2002.
\newblock The vol smile problem.
\newblock {\em Risk}, February, 61--65.

\bibitem[\protect\citename{Lucic, }2008]{Lucic2008}
Lucic, V. 2008 (July).
\newblock {\em Boundary Conditions for Computing Densities in Hybrid Models via
  PDE Methods}.
\newblock SSRN 1191962.

\bibitem[\protect\citename{Rannacher, }1984]{Rannacher:1984}
Rannacher, R. 1984.
\newblock Finite Element Solution of Diffusion Problems with Irregular Data.
\newblock {\em Numerische Mathematik}, {\bf 43}, 309--327.

\bibitem[\protect\citename{Strang, }1968]{Strang}
Strang, G. 1968.
\newblock On the construction and comparison of difference schemes.
\newblock {\em SIAM J. Numerical Analysis}, {\bf 5}, 509--517.

\bibitem[\protect\citename{Tavella \& Randall, }2000]{fdm2000}
Tavella, D., \& Randall, C. 2000.
\newblock {\em Pricing Financial Instruments. The Finite-Difference method.}
\newblock Wiley series in financial engineering.
\newblock New York,: John Wiley \& Sons, Inc.

\bibitem[\protect\citename{Toivanen, }2010]{toivanen2010}
Toivanen, J. 2010.
\newblock A Componentwise Splitting Method for Pricing American Options Under
  the Bates Model.
\newblock {\em Pages  213--227 of:} {\em Computational Methods in Applied
  Sciences}.
\newblock Springer.

\bibitem[\protect\citename{Wang {\em et~al.}, }2007]{WangWanForsyth2007}
Wang, I.R., Wan, J..W.L., \& Forsyth, P.~A. 2007.
\newblock Robust numerical valuation of European and American options under the
  CGMY process.
\newblock {\em J. Comp. Finance}, {\bf 4}, 31--70.

\end{thebibliography}
\newcommand{\noopsort}[1]{} \newcommand{\printfirst}[2]{#1}
  \newcommand{\singleletter}[1]{#1} \newcommand{\switchargs}[2]{#2#1}

\newpage
\appendix
\numberwithin{equation}{section}
\appendixpage
\section{Construction of the Backward Time Evolution Operator}
Here our goal is as follows. Starting with the finite-difference scheme given in \eqref{HV}, presented in the form of fractional steps, we wish to find an alternative representation of this scheme in the form of \eqref{BKE1}. In other words, given \eqref{HV}, we are looking for an explicit representation of the operator $\mathcal{B}$ in \eqref{BKE1}.

Recall \eqref{HV}:
\begin{align*}
Y_0 &= V_{n-1} + \Delta \tau F(\tau_{n-1},V_{n-1}), \\
Y_j &=  Y_{j-1} + \theta \Delta \tau \left[F_j(\tau_n,Y_j) - F_j(\tau_{n-1},V_{n-1})\right], \ j = 1, \dots, k \nn \\
\tilde{Y}_0 &= Y_0 + \dfrac{1}{2} \Delta \tau \left[F(\tau_n,Y_k) - F(\tau_{n-1},V_{n-1})\right], \nn \\
\tilde{Y}_j &= \tilde{Y}_{j-1} + \theta \Delta \tau \left[F_j(\tau_n,\tilde{Y}_j) - F_j(\tau_n,Y_k)\right], \ j=1,\dots,k \nn \\
V_n &= \tilde{Y}_k, \nn
\end{align*}
\noindent and proceed constructing the whole transition operator $R$ step by step along each line of \eqref{HV}. For the LSV problem, $k=2$.

Continuing, we will use the same notation for the transposed vectors $V_n$, $Y_k$, $\tilde{Y}_k$ as it should not bring any ambiguity. Also for convenience we denote $F_i^n = F_i(\tau_n)$. We can then write the first equation in \eqref{HV} as
\begin{equation} \label{1tr}
Y_0 = (I + \Delta \tau F^{n-1}) V_{n-1},
\end{equation}
where $F(\tau)$ is treated as an operator (or, given a finite-difference grid $\mathbf{G}$, the matrix of the corresponding discrete operator). We also reserve symbol $I$ for an identity operator (matrix).

It is important to notice that operators $F$ at every time step do not explicitly depend on a backward time $\tau$, but only via time-dependence of the model parameters.\footnote{We allow coefficients of the LSV model be time-dependent. However, they are assumed to be piece-wise constant at every time step.}

Proceeding in the same way, the second line of \eqref{HV} for $j=1$ is now
\[ (I - \theta\Delta \tau F^n_1) Y_1 = Y_0 - \theta \Delta \tau F^{n-1}_1  V_{n-1}.
\]
Therefore,
\begin{align} \label{2_1_tr}
Y_1 &= M_1^{-1}\left[Y_0 - \theta \Delta \tau F^{n-1}_1 V_{n-1}\right]
= M_1^{-1}\left[I + \Delta \tau \left(F^{n-1} -\theta F_1^{n-1}\right)\right]
V_{n-1}, \\
&M_i \equiv I - \theta\Delta\tau F^n_i. \nn
\end{align}
Similarly, for $j=2$ we have
\begin{align} \label{2_2_tr}
Y_2 &= M_2^{-1}\left[Y_1 - \theta \Delta \tau F^{n-1}_2 V_{n-1}\right] = R_2 V_{n-1}, \\
&R_2 \equiv M_2^{-1}\left\{ M_1^{-1}\left[ I + \Delta \tau \left(F^{n-1} - \theta F^{n-1}_1 \right)\right] - \theta \Delta \tau F^{n-1}_2\right\}. \nonumber
\end{align}
The third line in \eqref{HV} is now
\begin{equation} \label{3tr}
\tilde{Y_0} = Y_0 + \frac{1}{2}\Delta \tau \left(F^n Y_2 - F^{n-1} V_{n-1}\right) = \left[I + \frac{1}{2} \Delta \tau \left(F^{n-1} + F^n R_2\right) \right]V_{n-1}.
\end{equation}
The last line \eqref{HV} for $j=1,2$ can be transformed to
\begin{equation} \label{4_1_tr}
\tilde{Y_j} = M_j^{-1} \left(\tilde{Y}_{j-1} - \theta \Delta \tau F^n_j Y_2\right), \quad j=1,2.
\end{equation}
Collecting all lines together we obtain
\begin{align} \label{explR}
V_n &= R V_{n-1} \\
& R \equiv M_2^{-1}\left\{M_1^{-1}\left[I + \frac{1}{2} \Delta \tau \left(F^{n-1} + F^n R_2\right) - \theta \Delta \tau F^n_1 R_2\right] - \theta \Delta \tau F^n_2 R_2 \right\}. \nn
\end{align}

Thus, we managed to find an explicit representation for $R$ that follows from the finite-difference scheme in \eqref{HV}. To construct a transposed operator $R^T$, we use well-known rules of matrix algebra to get
\begin{align} \label{Rtrans}
R^T &= \left\{M_1^{-1}\left[I + \frac{1}{2} \Delta t \left(F^{n-1} + F^n R_2\right) - \theta \Delta t F^n_1 R_2\right] - \theta \Delta t F^n_2 R_2 \right\}^T (M_2^T)^{-1} \\
&= \left\{\left[I + \frac{1}{2} \Delta t \left((F^{n-1})^T + R^T_2 (F^n)^T \right) - \theta \Delta t R^T_2 (F^n_1)^T \right](M_1^T)^{-1} - \theta \Delta t R_2^T (F^n_2)^T \right\} (M_2^T)^{-1}, \nn \\
R_2^T &= \left\{\left[I + \Delta t \left((F^{n-1})^T - \theta (F^{n-1}_1)^T  \right) \right](M_1^T)^{-1} - \theta \Delta t (F_2^{n-1})^T \right\}(M_2^T)^{-1}, \nn
\end{align}
\noindent where we have assumed $\Delta \tau = \Delta t$.

\end{document}